\documentclass[preprint]{aastex}

\usepackage{graphicx}
\usepackage{amsmath,amssymb}
\usepackage[usenames,dvipsnames]{color}
\usepackage[colorlinks=true, citecolor=blue, linkcolor=WildStrawberry]{hyperref}
\usepackage{epstopdf}

\newcommand{\beq}{\begin{equation}}
\newcommand{\eeq}{\end{equation}}
\newcommand{\bdm}{\begin{displaymath}}
\newcommand{\edm}{\end{displaymath}}

\DeclareFontFamily{OT1}{pzc}{}
\DeclareFontShape{OT1}{pzc}{m}{it}{<-> s * [1.10] pzcmi7t}{}
\DeclareMathAlphabet{\mathpzc}{OT1}{pzc}{m}{it}

\graphicspath{{./plots/}}
\begin{document}

\title{A Daytime Measurement of the Lunar Contribution to\\ 
the Night Sky Brightness in LSST's {\it ugrizy} Bands-- Initial Results}

\author{Michael Coughlin}
\affil{Department of Physics, Harvard University, Cambridge, MA 02138, USA}

\author{Christopher Stubbs}
\affil{Department of Physics, Harvard University, Cambridge, MA 02138, USA\\
Department of Astronomy, Harvard University, Cambridge MA 02138, USA}

\author{Chuck Claver}
\affil{LSST Observatory, Tucson, AZ, 85712}

\begin{abstract}

We report measurements from which we determine the spatial structure of the lunar contribution to night sky brightness, taken at the LSST site on Cerro Pachon in Chile. We use an array of six photodiodes with filters that approximate the Large Synoptic Survey Telescope's {\it u, g, r, i, z,} and {\it y} bands. We use the sun as a proxy for the moon, and measure sky brightness as a function of zenith angle of the point on sky, zenith angle of the sun, and angular distance between the sun and the point on sky. We make a correction for the difference between the illumination spectrum of the sun and the moon. Since scattered sunlight totally dominates the daytime sky brightness, this technique allows us to cleanly determine the contribution to the (cloudless) night sky from backscattered moonlight, without contamination from other sources of night sky brightness. We estimate our uncertainty in the relative lunar night sky brightness vs. zenith and lunar angle to be 10\,\%. This information is useful in planning the optimal execution of the LSST survey, and perhaps for other astronomical observations as well. Although our primary objective is to map out the angular structure and spectrum of the scattered light from the atmosphere and particulates, we also make an estimate of the expected number of scattered lunar photons per pixel per second in LSST, and find values that are in overall agreement with previous estimates. 

\end{abstract}

\maketitle

\section{Introduction}
\label{sec:Intro}
Efficient telescope scheduling is essential to maximize the scientific output of survey telescopes. 
Optimizing a survey's scientific merit function requires scheduling decisions that consider slew rate, sky brightness, source location, transparency and extinction, and many other factors. For surveys such as LSST that will scan the entire accessible sky \citep{Ivezic2014}, a determination of the sky brightness as a function of azimuth, elevation and passband is an important factor in crafting an optimal sequence of observations. Contributions to the night sky include unresolved or diffuse celestial sources, emission from OH molecules in the upper atmosphere, zodiacal light, man-made light pollution, and moonlight that scatters from clouds and from the constituents of the atmosphere. Some of these contributions to the night sky are stable over time and do not impact the order in which we observe fields. Other 
contributions to the night sky are time-variable but deterministic. The monthly 
variation due to moonlight falls in this category, for cloud-free conditions. Other 
contributions to sky brightness, such as variable OH emission \citep{High2010} and moonlight scattering from clouds, are more stochastic in nature. 

Understanding the brightness of the cloudless moonlit sky in the LSST bands is one key component in scheduling decisions. To make predictions of potential future performance, LSST has developed an operations simulator to study different scheduling algorithms \citep{OpSim}. 
Thus far, the LSST operations simulator has used historical weather records and measures of the atmospheric conditions for the Cerro Pachon site (taken over a 10-year time period). 
The operations simulator generates a sky model that predicts sky brightness based on the Krisciunas and Schaefer model \citep{KrSc1991}, which is further discussed below. It also simulates atmospheric seeing and cloud coverage as a function of time. This information is used to estimate the efficiency of the survey for different candidate scheduling algorithms. As LSST advances into the construction phase, and eventually into full operation, we need a higher fidelity determination of the brightness of the cloudless 
night sky. This will be augmented with all-sky-camera data \citep{AllSkyCamera} to make real-time, condition-dependent adjustments to the sequencing of LSST observations.  

We define the lunar contribution to sky brightness as the difference between the observed sky brightness (in units of magnitudes per square arc sec) with the Moon above the horizon and the moonless sky brightness, for a given the phase of the lunar cycle. From the standpoint of scheduling decisions, what matters most is the {\it relative} lunar brightness variation across the accessible sky, and so our primary goal in this paper is to determine this spatial structure, using the sun as a proxy for the moon. This allows us to obtain high signal-to-noise data, without complications from other contributions to sky brightness. 

An {\it ab initio} computation of the lunar sky illumination is complicated due to multiple scattering effects. Sunlight reflects off the moon, and a portion of this light is scattered towards the Earth. This light impinging on the top of the atmosphere is then scattered by molecules and aerosols, and some is absorbed. The moonlight can be scattered multiple times, including off of the ground, before it reaches the telescope pupil. An empirical measurement is arguably more secure than a radiative transfer calculation that must make assumptions about the size, shape, and vertical distribution of aerosols. 

Walker developed \citep{Walker1987} a scattered moonlight model that included a table of sky brightness in five photometric bands, at five different moon phases. It did not account for the positions of the Moon or observation target, and was measured during solar minimum. Because of these shortcomings, it is not accurate enough for current and future telescope operations. Later, Krisciunas and Schaefer used an empirical fit to 33 observations taken in the V-band taken at the 2800\,m level of Mauna Kea, resulting in an accuracy between 8\% and 23\% if not near full Moon \citep{KrSc1991}. This model predicted the moonlight as a function of the Moon's phase, the zenith distance of the Moon, the zenith distance of the sky position, the angular separation of the Moon and sky position, and the band's atmospheric extinction coefficient. More recently, a spectroscopic extension of this model was used to fit sky brightness data from Cerro Paranal \citep{Noll2012}. This treatment includes all relevant components, such as scattered moonlight and starlight, zodiacal light, airglow line emission and continuum, scattering and absorption within the Earth's atmosphere, and thermal emission from the atmosphere and telescope. This model was recently updated with an observed solar spectrum, a lunar albedo fit, and scattering and absorption calculations \citep{Jones2013}. Winkler et al. characterized the nighttime sky brightness profile under a variety of atmospheric conditions using measurements from the South African Astronomical Observatory soon after the Mount Pinatubo volcanic eruption in 1991 \citep{WiWy2013}.
Our goals in this paper are more limited, as we are primarily interested in the 
spatial structure and spectrum of the scattered moonlight component of the night sky. 

Based on this discussion, it is clear that models for scattered moonlight are very complicated. This motivates our attempt to empirically determine the relative sky brightness as a function of lunar phase, and its dependence on the positions of the target and the moon. We measured the solar sky brightness as a function of angle between sky location and the sun, as well as zenith angles of the sun and the telescope. We argue that this is useful because up to wavelength-dependent lunar albedo factors, the sky illumination pattern that the moon casts has the same functional form as from the sun in the daytime. 

We have made measurements of the daytime sky brightness at Cerro Pachon, the LSST site in Chile, with an array of six photodiodes with filters in 
the {\it u, g, r, i, z,} and {\it y} bands. 
There is an extensive history of daytime sky brightness measurements. In particular, the 
angular and wavelength dependence of the observed solar scattering can be used to deduce properties of atmospheric aerosols and precipitable water vapor. 
We use a similar measurement scheme to the AERONET remote sensing aerosol monitoring network \citep{Holben1998}. The AERONET sky brightness data are taken in optical passbands that differ from those that LSST will use. Rather than invoke a set of 
color transformations to convert from AERONET into LSST bands, here we make a 
direct measurement of sky brightness in the LSST passbands. 

We describe the apparatus in section \ref{sec:apparatus}. Measurements and analysis are presented in section \ref{sec:results}. We conclude with a discussion of topics for further study in section \ref{sec:Conclusion}. 

\section{Apparatus}
\label{sec:apparatus}

\begin{figure}[t]
 \includegraphics[width=6.0in]{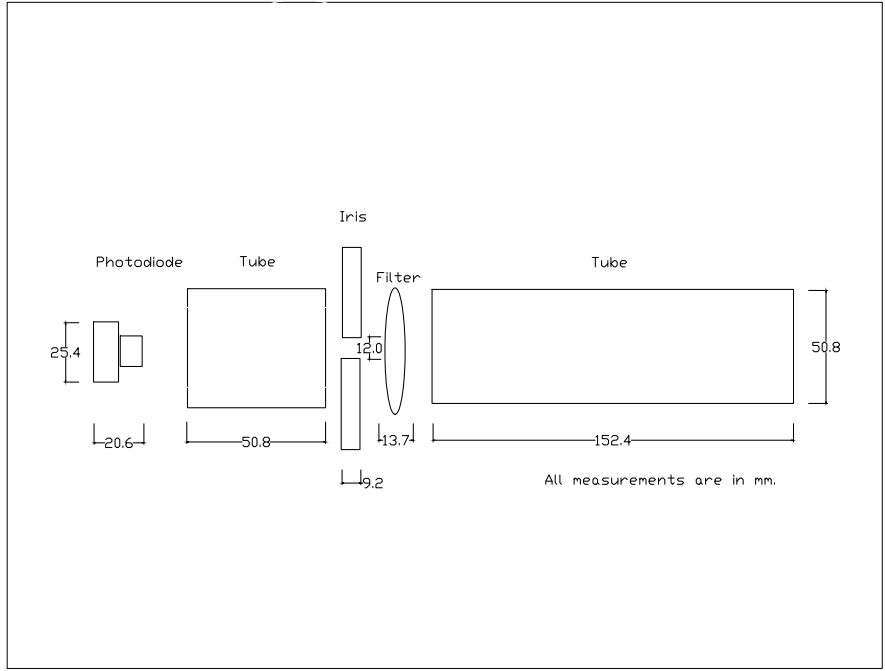}
 \caption{Sketch of the photodiode portion of the apparatus. The photodiode mounts include a photodiode, an iris, a filter, and a baffle tube.}%
 \label{fig:apparatus}
\end{figure}

Fig.~\ref{fig:apparatus} shows a sketch of the photodiode mount, which is identical 
for all six channels except for the interference filters that define the passbands. 
Light enters a 50\,mm inner diameter cylinder, 152.4 mm long, that serves to block off-axis stray light. The 50\,mm diameter filters are placed at the base of this baffle tube. 

\begin{figure}[htbp]
\begin{center}
\includegraphics[width=6.0in]{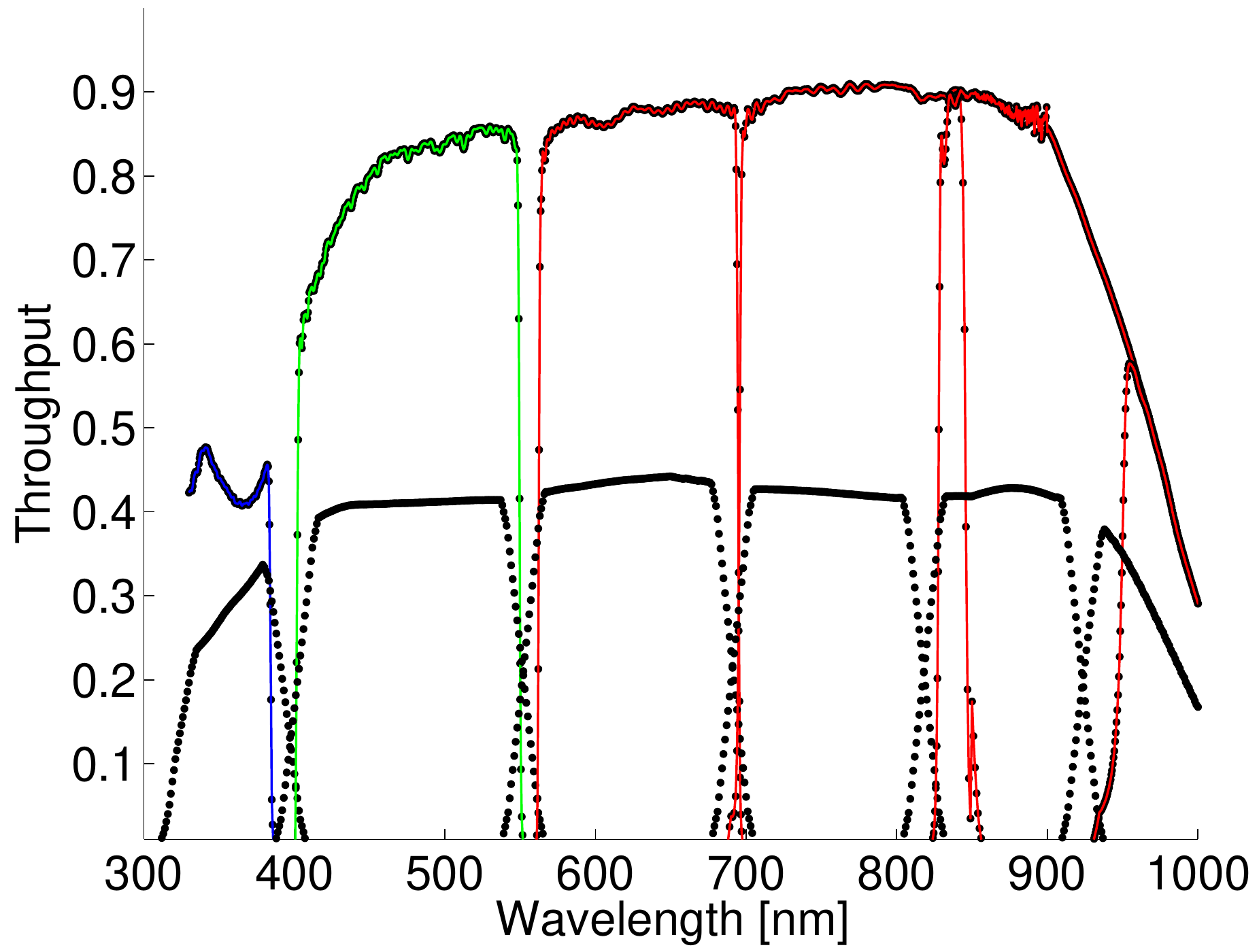}
\caption{Photon sensitivity function curves for the photodiodes (upper) used in this experiment, and 
the expected photon sensitivity function for LSST (lower curves, due to more complex optical system). From left to right the bands are $u,g,r,i,z$ and $y$. We use this information
to make a throughput correction when predicting scattered moonlight backgrounds for LSST. The Astrodon filters we used for the photodiodes are designed to avoid the water band at 940 nm. }
\label{fig:QE}
\end{center}
\end{figure}

We used ``Generation 2 Sloan Digital Sky Survey (SDSS)'' {\it u,g,r,i,z,y} filters from Astrodon \citep{Astrodon}. 
Figure~\ref{fig:QE} shows their transmission spectra as well as the current-design LSST filters \citep{Ivezic2014}, for comparison. The Astrodon filters we used are essentially flat-topped, with minimal leakage or in-band ripple. An adjustable iris (Thor Labs SMD12C) sits behind the interference filter. We found we could operate with these irises set to their maximum opening diameter of 12\,mm, for all six passbands. 
The only other transmissive optical element that lies between the Si and the sky is a quartz window in front of the photodiode. 

The photodiodes are SM1PD2A cathode-grounded Si UV-enhanced photodiodes, obtained  from Thor Labs. The photodiodes have a 10\,mm x 10\,mm active area behind a 9\,mm diameter input aperture. The only other transmissive optical element that lies between the Si and the sky is a quartz window in front of the photodiode. The etendue of the system is established by the combination of the 12\,mm diameter iris and the 9\,mm circular photodiode input aperture. These two circular apertures are separated by a distance of 60\,mm. 

\subsection{Etendue of the photodiode plus tube system, in comparison to an LSST pixel}

As seen from the plane of the diode aperture, the full-angle subtended by the adjustable iris is then $2\,\textrm{arctan}(\frac{6}{60})=11.4^\circ$, which subtends a solid angle of 
$\Omega_\textrm{diode}=2\,\pi\,(1-\cos(\frac{11.4}{2}))=3.11\times10^{-2}$ steradians. For comparison, 
a pixel on LSST subtends 0.2 arcsec on a side, for a solid angle of $7.4\times10^{-13}$\,steradians/pixel.

If we consider the iris as establishing the field of view of the photodiode system, then the 
aperture in front of the diode determines the sensor's unvignetted collecting area, where $A_\textrm{photodiode}=\pi\,(4.5 \times 10^{-3} \textrm{m})^2 = 6.36\times10^{-5}\,\textrm{m}^2$. This amounts to computing the overlap of the sensor and iris apertures. For comparison, the effective collection area of LSST is equivalent to a diameter of 6.5\,m, for a collection area of $A_\textrm{LSST}=\pi\,(\frac{6.5\textrm{m}}{2})^2=33.2\,\textrm{m}^2$. The ratio of the etendue of an LSST pixel to the photodiode is then $R=\frac{A_\textrm{LSST} \Omega_\textrm{LSST pixel}}{A_\textrm{photodiode} \Omega_\textrm{photodiode}}=1.23\times 10^{-5}$.

%


The interpretation of the data will benefit from knowing the ratio between the instrumental response function of the photodiode system and LSST. Table~\ref{tab:throughputs} compares the band-integrated system throughput figure for the photodiode system (the Thor labs QE times the Astrodon filter response)  with two versions of the LSST throughput. LSST is considering using CCDs from  two vendors, e2v and ITL, and these have somewhat different quantum efficiency curves. We have, therefore, provided in Table~\ref{tab:throughputs} the results from integrating over the response functions (including in the LSST case the three reflections, the obscuration, the filter and corrector transmissions, and the detector QE) at a spacing of one nm. The units in Table~\ref{tab:throughputs} are nm, and can be interpreted as the sensitivity-weighted equivalent width of the respective filters. Taking the ratio of these numbers, passband by passband, allows us to scale the photodiode measurements to anticipated values on the LSST focal plane. 

\begin{table}[htdp]
\begin{center}
\begin{tabular}{|c|c|c|c|c|c|}
\hline
band & diodes & LSST with ITL & LSST with e2v & $\left [   \frac{T_{ITL}} {T_{diodes}}  \right ]$ & $\left [   \frac{T_{e2v}} {T_{diodes}}  \right ]$ \\
\hline
u & 33.8 & 20.6 & 15.3 & 0.61 & 0.45  \\
g & 99.0 & 61.3 & 65.4 & 0.62 & 0.66 \\
r & 93.2 & 60.3 & 62.9 &0.65 & 0.67 \\
i & 106.7 & 53.7 & 53.2 & 0.50 & 0.50 \\
z & 155.2 & - & - & - & - \\
zs & 65.8 & 44.3 & 43.3 & 0.67 & 0.66  \\
y & 69.5 & 27.9 & 27.2 & 0.40 & 0.40 \\
\hline
\end{tabular}
\end{center}
\caption{System throughput values. The first three columns are the integral of the system response function at 1 nm spacings, for each passband in the different systems. 
The last two columns show ratios of the LSST
throughput to that of the diodes. The diode instrument has no reflective optics 
and a minimum of air-glass interfaces, whereas LSST has three reflections from 
aluminum as well as a three element corrector. Also, the photodiodes are considerably thicker than the LSST CCDs, and have enhanced UV sensitivity. This accounts for the 
increased diode throughput compared to the LSST system.}
\label{tab:throughputs}
\end{table}%

%
%
%
%

The photodiodes are connected via coaxial cable to a set of manual switches that feed a selected one of the six signals to a Thor Labs model PDA200C photocurrent amplifier, which produces a $\pm$ 10\,V signal proportional to the current from the selected photodiode. This signal was connected to an Arduino Uno, which digitizes this signal with a 10 bit A/D converter. The Arduino was connected to a serial port on the data collection computer.

The six photodiode tubes are mounted on a Celestron model CG-5 equatorial telescope mount, which is controlled by external connection to a laptop computer. Stellarium (\cite{Stellarium}) is used to control the telescope mount pointing. The mount's RA and DEC motors have a precision of 0.05$^\circ$.  The computer registered the right ascension ($\alpha$) and declination ($\delta$) for each brightness measurement, along with the photocurrent from each of the six photodiodes. 
These $\alpha$ and $\delta$ measurements were converted to alt-az coordinates for data analysis, since as shown below this is the most natural angular coordinate system for this problem. 

%
%

\subsection{Sky Scanning Strategy and Angular Coordinates}

Assuming that the scattering properties of the atmosphere are axisymmetric about local vertical, the normalized sky brightness (scaled to the brightness of the illuminating source) depends on three angles: the zenith angle $z_\textrm{source}$ of the source (sun or moon), the zenith angle $z_\textrm{tel}$ of the telescope boresight, and the azimuthal angle $\Delta \phi$ between the source and the boresight.
    
An ``almucantar'' is the line on the sky at the elevation angle of the Sun, at some
given time. An advantage to making sky brightness measurements along an almucantar 
is that the boresight and source elevation angles are constant, and equal. Only their
azimuthal separation is varied. 
During an almucantar measurement, observations are made at the solar elevation angle through 360$^\circ$ of azimuth. The almucantar sweep is a special case of a constant-zenith-angle scan, which is our favored data collection method. The range of scattering angles along an almucantar decreases as the solar zenith angle decreases; thus almucantar sequences made at airmass of 2 or more achieve maximum scattering angles of 120$^\circ$ or larger.

We elected to obtain our sky brightness data in a succession of constant-zenith angle scans, taking a data point every 45$^\circ$ of azimuth, except near the zenith. This gives us 8 data points in azimuth at each telescope zenith angle. We obtained data  at zenith angles of 0, 30, 45, 60, and 75$^\circ$, and along the almucantar, over the course of the day, in each passband. The resulting daytime sky brightness (DSB) data by passband, $DSB(z_\textrm{source}, z_\textrm{tel}, \phi, \textrm{filter})$, comprise our measurement. We generate an all-sky map of sky brightness. The data are processed as follows. For each solar zenith angle, we generate an all-sky map of brightness vs. position. We then repeat for different values of solar zenith angle. We make a polynomial fit to brightness vs. altitude and delta-azimuth. This is a map of relative night sky brightness if the moon were at the location of the sun, up to an overall scale factor per passband. We can use the geometry of the photodiode tube to compute number of photons per square arcsec per square meter and then scale the value by about 14 magnitudes to get the lunar contribution. The scaling factors are computed in the next section. We can then compute an estimate for other lunar phases, based on the lunar phase function. Of course, the actual sky brightness is a combination of the lunar contribution, which we compute, plus other factors, which depend on the instrument, plate scale, integration time, etc. as well as solar cycle and site characteristics. 
  
\subsection{Scaling from Solar to Lunar Illumination}

Keiffer and Stone (\cite{KiSt2005}, hereafter K\&S)) describe how to scale between solar 
illumination and lunar
illumination at the top of the atmosphere, depending on both reflection geometry
and wavelength. The ratio $R$ of the lunar to solar irradiance at the top of the atmosphere 
is given by
$R(\lambda,g)=A(\lambda, g) \left [ \Omega_M/\pi \right ] $,  where $\Omega_M$ 
is the solid angle subtended by the moon, and $A(\lambda,g)$ is a wavelength 
dependent scattering function that depends on the angle, g, between the vectors 
from the moon to the Earth and the moon to the sun. We
have assumed nominal values for the sun-moon and moon-earth distances. 
The geometrical dilution factor is 6.42/$\pi \times 10^{-5} = 2.04 \times 10^{-5}$, or 11.72 magnitudes. 

To correct for the wavelength-dependent and phase-angle-dependent lunar scattering function, we took the parametric description of $A(\lambda,g)$ provided in K\&S, integrated across our passbands (note: truncated u band at 350, no data bluer), to determine the scattering-dependent magnitude differences between sunlight and moonlight at the top of the atmosphere, for different lunar phases. In order to avoid the sharp peak in reflection due
to the ``opposition effect'', we limited the range of phase angles to $|g| > $ 2 degrees.

The additional attenuation from lunar scattering as a function of passband at 
full moon (taken here to be our minimum scattering
angle of 2 degrees) is listed in Table \ref{tab:fullmoon}. We obtained these values by numerically computing 

\begin{equation}
\Delta \textrm{mag}_i (g) = -2.5 \textrm{log}_{10} \left( \frac{\int A(\lambda,g) T_i(\lambda) {\rm d \lambda}} {\int T(\lambda) {\rm d \lambda}} \right)
\label{eq:fullmoon}
\end{equation}

\noindent
where $T_i (\lambda)$ is a top-hat approximation of filter $i$. 

\begin{table*}
\begin{center}
\begin{tabular}{|c|c|c|c|c|}
\hline
Band & $\Delta \textrm{mag}(g=2^\circ)$ & $\Delta \textrm{mag}(g=10^\circ)$ & $\Delta \textrm{mag}(g=45^\circ)$ & $\Delta \textrm{mag}(g=90^\circ)$  \\
\hline
$u$ & 2.60 & 3.05 &  4.06 & 5.52\\
$g$ & 2.36 & 2.78 & 3.77 & 5.19 \\
$r$ &  2.10 & 2.50 & 3.45 & 4.84 \\
$i$ & 1.92 & 2.31 & 3.23 & 4.59 \\
$z$ & 2.17 & 2.57 & 3.53 & 4.92 \\
$y$ & 1.72 & 2.12 & 2.91 & 4.31 \\
\hline
\end{tabular}
\end{center}
\caption{Irradiance attenuation due to lunar scattering, in the LSST bands, at various lunar phase angles. The illumination from the moon is slightly redder than sunlight, in general. 
This reddening effect increases as the phase angle increases.}
\label{tab:fullmoon}
\end{table*}%

Based on these values, since the sky brightness scales linearly with the irradiance provided at the top of the atmosphere, we expect the $r$ band full-moon lunar sky brightness to be a factor of 11.72+2.10=13.82 magnitudes fainter than what we observe in the daytime, if the moon were placed in the same alt-az position as the sun. 

%

This allows us to generate, up to a single passband-dependent overall scale factor that depends on the 
effective etendue of the photodiode tube, the equivalent full-moon sky brightness map
for the case where the moon is in the same location as the sun. 
We simply take the solar-illuminated sky brightness data, and scale all values to the
r band brightness at the zenith. Then we make a passband-dependent adjustment based
on the color of the reflected sunlight as shown in Table \ref{tab:fullmoon}.

\section{Results}
\label{sec:results}

\begin{table}[t]
\begin{center}
\begin{tabular}{|c|c|c|}
\hline
Filter & Dark Current [nA] & $1 \sigma$\,Uncertainty [nA] \\\hline\hline
u & 4.3 & 0.2 \\\hline
g & 2.1 & 0.3 \\\hline
r & 5.3 & 0.2 \\\hline
i & 4.2 & 0.2 \\\hline
z & 4.6 & 0.2 \\\hline
y & 2.0 & 0.3 \\\hline
\end{tabular}
\end{center}
\caption{Dark current measurement for all six filters. Measurements were taken periodically during data taking to check if there was a significant temperature dependence. Dark current values were found to be approximately the same over the run, and these values are about 100 times smaller than the currents in the sky brightness analysis.
}
\label{tab:DarkCurrent}
\end{table}

We begin by measuring dark current values for each photodiode channel, and the results are displayed in Table~\ref{tab:DarkCurrent}. The photodiodes have dark current values ranging from 2.0 to 5.3\,nA with statistical uncertainties between 0.2 to 0.3\,nA. These dark currents are well under 1\% of the signal levels from the sky. Because it is a negligible contribution, we ignore the dark current contribution in the analysis that follows.

\subsection{Observations}

We obtained sky brightness data from the roof of the ALO building on Cerro
Pachon, (located at S 30:15:06, W 70:44:18) during the daytime on 
2014 Sept 4, 5, 6 and 7. The conditions on Sept 5 were less favorable, 
with high cirrus clouds in the sky. 
We cycled through the sky sampling strategy described above, 
taking 2000 data points in each passband, running through the 6 bands
in succession. Each data collection period at a fixed pointing lasted about 
two minutes, and a full cycle across the sky lasted about an hour. In all, we collected 10 sequences, spanning a range of solar elevations from 20$^\circ$ to 55$^\circ$. 

\subsection{Spatial structure of scattered light}

Night sky structure was investigated by \cite{ChHa1996}, in the context of flat-fielding. The authors are unaware of a comprehensive program to map (and visualize) the sky brightness under variable lunar illumination conditions. In the following, we show the sky brightness dependence on the zenith angle $z_\textrm{source}$ of the source (sun or moon), the zenith angle $z_\textrm{tel}$ of the telescope boresight, and the azimuthal angle $\phi$ between the source and the boresight. We perform fits of sky brightness to the three independent variables and compute the effect on 5 sigma point source detection magnitude for a survey such as LSST. The data allows us to study the color across the sky. We also note how to scale overall brightness in each band as a function of lunar phase.

\subsection{Spatial Structure of Lunar Sky Brightness}

\begin{table*}
\begin{center}
\begin{tabular}{|c|c|c|c|c|c|}
\hline
Band & a ($\times 10^{11}$) & b ($\times 10^{11}$) & c ($\times 10^{11}$) & d ($\times 10^{11}$) & Median Residual ($\times 10^{11}$) \\
\hline
u & 88.5 (6.2) & -0.5 (0.1) & -0.5 (0.1) & 0.4 (0.1) & 5\\
g & 386.5 (34.0) & -2.2 (0.2) & -2.4 (0.2) & 0.8 (0.5) & 13\\
r & 189.0 (32.7) & -1.4 (0.2) & -1.1 (0.2) & 0.8 (0.5) & 11\\
i & 164.8 (33.1) & -1.5 (0.2) & -0.7 (0.2) & 0.6 (0.5) & 12\\
z & 231.2 (62.3) & -2.8 (0.3) & -0.7 (0.4) & 1.4 (0.9) & 21\\
zs & 131.1 (45.6) & -1.4 (0.2) & -0.5 (0.3) & 0.2 (0.6) & 10\\
y & 92.0 (32.7) & -1.3 (0.2) & -0.2 (0.2) & 0.9 (0.5) & 20\\\hline
\end{tabular}
\end{center}
\caption{Daytime sky brightness values as function of angle between the point on sky and sun, altitude of the point on the sky, and altitude of sun fit to a plane of the form $a + b x + c y + d y$ in electrons/s. The zs band values, which are given to approximate the LSST z filter, are computed using Astrodon z minus y. Here, x corresponds to the angle between the point on the sky and the sun, y corresponds to the altitude of the point on the sky, and z corresponds to the altitude of the sun. The median of the absolute value of the residuals is given by the final column. }
\label{tab:results}
\end{table*}

The first result we present is sky brightness as a function of angle between the point on sky and the sun, the altitude of the point on the sky, and the altitude of the sun. The measurements describe a three-dimensional surface corresponding to these parameters. We convert the photodiode output from $\mu$A to electrons/s. We fit resulting data to a plane of the form $a + b x + c y + d y$ for each of the six bands. Here, x corresponds to the angle between the point on sky and the sun, y corresponds to the altitude of the point on the sky, and z corresponds to the altitude of the sun. The coefficients and the standard errors, as well as the median of the residuals for the fits are shown in Table~\ref{tab:results}. We find the residuals of the fits to be small; these are generally an order of magnitude smaller than the overall scale factor (a). There are a number of notable features. The first is that as $\phi$ increases, the sky brightness decreases. This in and of itself is not surprising, but the rates of decrease are significant. For example, in the g band, for every 10 degrees that $\phi$ decreases, the number of photons by more than a factor of 2. There is a similar effect for the altitude of the point on the sky. As the point moves to the horizon, the sky brightness increases. This effect is more pronounced for the bands near the blue. Finally, the sky brightness increases as the altitude of the sun increases. Perhaps more interesting is the significant color dependence of the results. Although the trends described above hold true regardless of color, the magnitude of their effect is very different. The difference between the g and y bands is about a factor of 2 difference in the point on the sky dependence and more than a factor of 10 difference in the altitude of the point on the sky. The effect of the altitude of the sun, $z_\textrm{source}$, is more constant across color.

It is straightforward to apply the measured planar fit coefficients to an individual observation. Table~\ref{tab:fullmoon} provides the appropriate scale factor for different moon phases and passbands. If the passband of interest varies significantly from the filters in this study, one can compute the appropriate factor from equation~\ref{eq:fullmoon}. After this, one finds the altitude and azimuth at the site of interest at the time of the observation as well as the altitude and azimuth of the target. Three angles are then computed from these quantities: the angle between the point on the sky and the moon, the altitude of the point on the sky, and the altitude of the moon. One then computes $a + b x + c y + d y$ for the appropriate passband, where a, b, c, and d are given in Table~\ref{tab:results}, and x corresponds to the angle between the point on the sky and the moon, y corresponds to the altitude of the point on the sky, and z corresponds to the altitude of the moon.
This table allows for a straightforward comparison between the ratio of sky brightnesses for different colors. The ratio of fluxes in u to g, for example, are fairly flat across the sky due to the similar ratios between coefficients. On the other hand, the ratio of fluxes in u to i, depends heavily on angle to the source, with virtually no dependence on zenith angle. This indicates that the sky is much redder close to the moon than far away.
A code that performs these steps is available at https://github.com/mcoughlin/skybrightness for public download. Hopefully, this will allow other researchers to easily use the data product. Required inputs are the latitude, longitude, and elevation of the site, right ascension and declination of the source, the passband of interest, and the times of observation.



\subsection{From Relative Sky Brightness to $m_5$ Variations to Optimal Scheduling}

\begin{figure}[htbp]
\begin{center}
\includegraphics[width=5.5in]{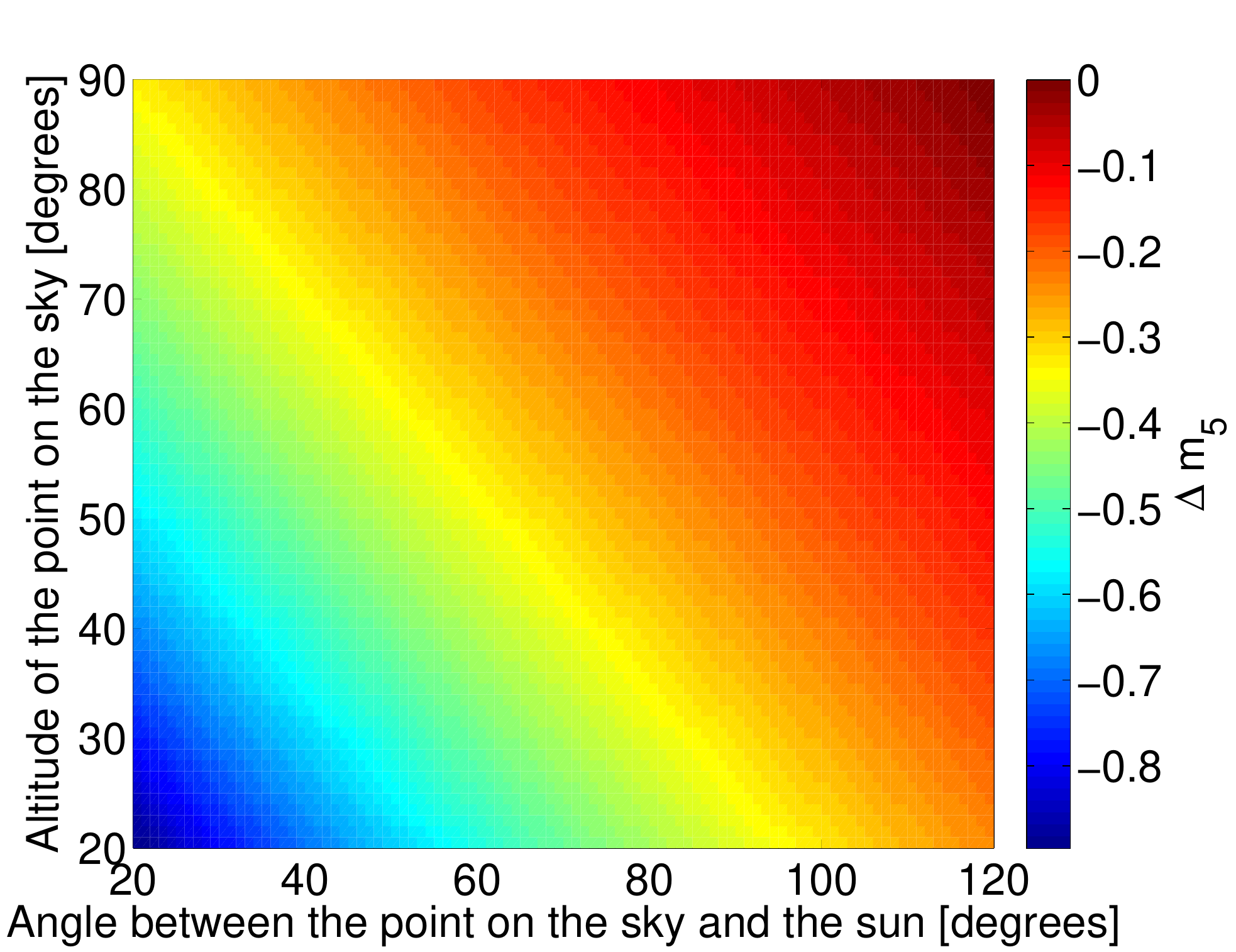}
\caption{Variation $\Delta m_5$ in the point source magnitude that can be detected at 5$\sigma$ in the u-band, against spatially varying sky brightness. This contour plot shows the change in point 
source detection threshold as a function of altitude and angular 
separation from the moon.
The color bar indicates the change in $m_5$ for a fixed exposure time, in magnitudes. Maps
such as this 
can be used to optimize the sequence of LSST observations. }
\label{fig:dm5}
\end{center}
\end{figure}

We can use the data we have generated of {\it relative} daytime sky brightness to generate a sky map of degradation in the point source magnitude that can be detected in the case where scattered moonlight dominates the Poisson noise. We define the {\it sky brightness factor}, SBF, to be the ratio of the local sky to the darkest attainable sky surface brightness at that moment. In order to achieve the same SNR with varying sky backgrounds, the source must be brighter by a factor of $\Delta m_5 = \frac{1}{2} \times -2.5\,\textrm{log}_{\textrm{10}}(SBF)$. Because the sky brightness structure is linear in the illumination level, the sun-illuminated measurements are perfectly valid for making these $\Delta m_5$ maps. If one region of the sky is twice as bright as another in the daytime, then for the moonlight-dominated case, replacing the sun with the moon will not change that fact. An example is shown in Figure \ref{fig:dm5}, for the u-band. For the scheduling of LSST observations over the course of a night, a week, or a month, this
is the format in which we think the sky brightness data is most useful. We stress that this comes directly from the daytime measurements of relative brightness, with no 
conversion needed, as long as scattered moonlight dominates the sky background. 

\subsection{Prediction of Scattered Moonlight Contribution to LSST backgrounds}

Table~\ref{tab:zenithresults} presents the data analysis sequence, for sky brightness obtained at zenith with a source elevation angle of 45$^\circ$. The table shows, for each passband, the measured photocurrent, the dark current value, the number of photoelectrons per second 
produced in the photodiode, the geometrical factor $GF$ for scaling from solar to lunar irradiance, the 
attenuation due to the lunar phase function at $PF$ full moon ($g=2^\circ$), the ratio $R$ of LSST pixel to photodiode etendues, the ratio of throughput times etendue for the two systems, and the
number of LSST photoelectrons per pixel per second. We compute

\begin{equation}
\Phi_\textrm{LSST}=\left [ \frac{I_\textrm{meas}-I_\textrm{dark}}{1.60 \times 10^{-19} Coul} \right ] * GF * PF * \left [   \frac{T_{LSST}} {T_{diodes}}  \right ] * \left [ \frac{(A \Omega)_{LSST~pixel}} {(A \Omega)_{diodes}} \right ]
\end{equation}
to obtain the expected number of photoelectrons per pixel per second we expect on the LSST focal plane, for full moon conditions, with the telescope pointed to the zenith, and a 
lunar zenith angle of 45$^\circ$.  Note that we do not include any factor for atmospheric attenuation since we wish to use the photon arrival rate on the photodiode to predict the lunar 
background flux on the LSST focal plane. 

\begin{table}[htdp]
\begin{center}
\begin{tabular}{|l|l|l|l|l|l|l|l|l|}
\hline
Band & I$_\textrm{meas}$ & $\Phi_\textrm{diode}$  & Geometry & Phase &  $\frac{T(ITL,e2v)\,A\,\Omega(LSST)}{TA\,\Omega(Photodiode)}$ & $\Phi_\textrm{ITL}$ & $\Phi_\textrm{e2v}$ & ETC \\
~ & $\mu$A & e/s & Factor & Factor & ~ & e/pix/s & e/pix/s &e/pix/s \\
\hline
u & 1.0  & $6.25\times 10^{12}$ & $2.04\times 10^{-5}$ & 0.091 & (0.61,0.45)*$1.23\times 10^{-5}$ & 86 & 64 & 106 \\
g & 3.5  & $6.25\times 10^{12}$ & $2.04\times 10^{-5}$ & 0.114 & (0.62,0.66)*$1.23\times 10^{-5}$ & 307 & 327 & 451 \\
r & 1.8  & $6.25\times 10^{12}$ & $2.04\times 10^{-5}$ & 0.14 & (0.65,0.67)*$1.23\times 10^{-5}$ & 168 & 173 & 186 \\
i & 1.5  & $6.25\times 10^{12}$ & $2.04\times 10^{-5}$ & 0.17 & (0.50,0.50)*$1.23\times 10^{-5}$ & 106  & 106 & 116 \\
z & 2.2  & $6.25\times 10^{12}$ & $2.04\times 10^{-5}$ & 0.13 & (0.67,0.66)*$1.23\times 10^{-5}$ & 210  & 207 & - \\ 
zs & 0.9  & $6.25\times 10^{12}$ & $2.04\times 10^{-5}$ & 0.13 & (0.67,0.66)*$1.23\times 10^{-5}$ & 84  & 83 & 89 \\
y & 1.1  & $6.25\times 10^{12}$ & $2.04\times 10^{-5}$ & 0.20 & (0.40,0.40)*$1.23\times 10^{-5}$ & 60 & 60 & 23 \\
\hline
\end{tabular}
\end{center}
\caption{Zenith daytime sky brightness values, converted to expected LSST 
lunar sky backgrounds at zenith at full moon. These are all adjusted to a common angle between the point on sky and sun, altitude of the point on the sky, and altitude of sun of 45$^\circ$. The zs band values, which are given to approximate the LSST z filter, are computed using Astrodon z minus y. There is qualitative agreement between the values calculated from the daytime measurement and the LSST exposure time calculator. The exposure time calculator uses the full sky spectrum, not just lunar part, and thus we expect the photodiode measurements to generally underestimate the exposure time calculator numbers.}
\label{tab:zenithresults}
\end{table}%
  


 

\section{Conclusion}
\label{sec:Conclusion}

Measurements of sky brightness are important for efficient telescope scheduling and predictions for LSST. While daytime measurements are approximations to Lunar measurements, it provides high signal to noise ratio measurements in the LSST bands. We measured the fall-off in sky brightness with angle from the sun and zenith angle.

There are a number of important conclusions to draw from the measurements. The first is that there are substantial gradients in scattered sky brightness, as much as 2 magnitudes. The second is that the scattered sky brightness increases closer to the horizon, perhaps due to 
more column density winning over extinction. The third is that when observing in bright time, there is significant benefit to point away from the moon.
 
\begin{figure}[htbp]
\begin{center}
\includegraphics[width=5.5in]{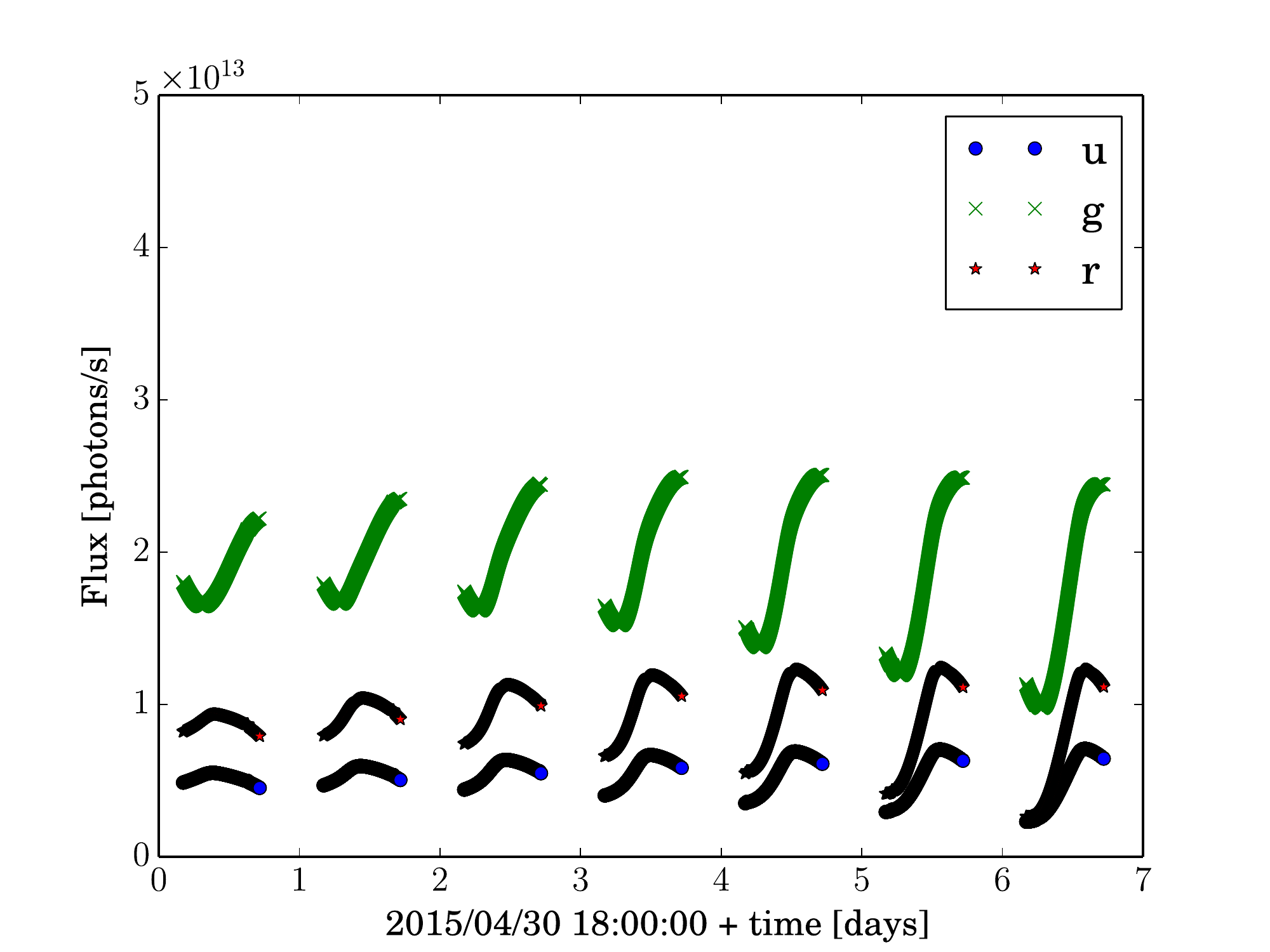}
\caption{u, g, and r-band flux as a function of time for a single point on the sky using data from Table~\ref{tab:zenithresults}. Code to produce this plot is available at https://github.com/mcoughlin/skybrightness for public download. This shows in general the differences in flux for a single observation throughout the night.}
\label{fig:deltam}
\end{center}
\end{figure} 
 
As motivation for future work, figure~\ref{fig:deltam} shows u, g, and r-band flux as a function of time for a single point on the sky using the measurements described in this paper.
We can use these measurements to prioritize observations throughout a given night.
In the future, we intend to improve on these measurements by designing the apparatus to take simultaneous measurements. 
With such an apparatus, we will be able to, for example, measure the color of clouds. 
We will also be able to take significantly more observations, allowing for refinement of this model, continuing to make it more useful for observers and those exploring scheduling strategies.

\section{Acknowledgments}
MC was supported by the National Science Foundation Graduate Research Fellowship
Program, under NSF grant number DGE 1144152. CWS is grateful to the DOE Office 
of Science for their support under award DE-SC0007881. Thanks also to Prof. Gary Swensen of Univ. of Illinois for hospitality in the ALO building at Pachon.

\bibliographystyle{plainnat}
\bibliography{references}

\end{document}